\documentstyle[12pt]{article}
\def\ni{\noindent} 
\def\Dc{{\cal D}}
\def\no{\nonumber}
\begin{document}
\begin{center}
{\Large {\bf Triplectic Quantization of W2 gravity}}\\
\vspace{1cm} 
%\end{center}

{\large Nelson R. F. Braga  and Cresus F. L. Godinho}  \\
\vspace{1cm}

 Instituto de F\'\i sica, Universidade Federal  do Rio de Janeiro,\\
Caixa Postal 68528, 21945  Rio de Janeiro,
RJ, Brazil\\

\vspace{1cm}

\end{center}

\vspace{1cm}
\abstract  
The role of one loop order corrections in the triplectic quantization
is discussed in the case of  W2 theory. This model illustrates the 
presence of anomalies and Wess Zumino terms in this quantization 
scheme where extended BRST invariance is represented  in a 
completely anticanonical form.


\vspace{4cm}

\noindent braga@if.ufrj.br; godinho@if.ufrj.br

\vfill\eject
\section{Introduction}
The so called triplectic quantization\cite{T1,T2,T3} is a 
general Lagrangian gauge theory quantization procedure following
the general lines of the field antifield or 
Batalin Vilkovisky (BV) method\cite{BV1,BV2}
but with the requirement of extended BRST\cite{He1,He2,He3}
(BRST plus anti-BRST) invariance rather than just BRST. 
In the usual BV quantization the BRST invariance is translated 
into the so called master equation. At zero loop order this 
equation is well defined and its solution, together with the 
appropriate requirements corresponding to gauge fixing, 
leads to the construction of the complete structure of ghosts, 
antighosts, ghosts for ghosts, etc\cite{HT,H1}. 
At higher orders in $\hbar$ one needs however to introduce some 
regularization procedure in order to give  a well defined meaning 
to the mathematical objects involved in the formal master
equation. Anomalies and Wess Zumino terms can this way be 
calculated at one loop order\cite{TPN,DJ}. 

In the triplectic quantization the extended BRST invariance
is translated into a set of two master equations 
corresponding to the requirements of BRST and 
anti-BRST invariances respectively. As in the standard BV case, both 
equations have formally an expansion in loop order. 
One then expects that anomalies and Wess Zumino terms should  show up 
at one loop order as long as one is able to introduce appropriate 
regularization schemes.
These features are not present in the recently discussed case of 
Yang Mills theory\cite{ABG}.  In that case only the zero loop order 
corrections are relevant, as there are no anomalies. 
The important features of calculation of anomalies and counterterms 
in the triplectic context have not yet been discussed in the literature.  
In this article we will discuss the W2 model where the one loop order 
corrections will nicely illustrate the behavior of the quantum master
equations, compared with the standard BV case. We will also show how 
to fix the gauge by means of canonical transformations.

\section{Triplectic quantization}
Considering some gauge theory, we enlarge the original field content 
$\,\phi^i\,$, adding all the usual gauge fixing structure: ghosts, 
antighosts and auxiliary fields associated with the original gauge 
symmetries. The resulting set will be denoted as $\,\phi^A\,$. 
Then we associate with each of these fields five new quantities, 
introducing the sets:  $ {\bar \phi}^A $, $\,\phi_A^{\ast\,1}\,$ ,
$\,\phi_A^{\ast\,2}\,$,$\,\pi_A^{1}\,$ and  $\pi_A^{2}\,$. 
The Grassmanian parities of these fields are: 
$\epsilon (\phi^A) \,=\,\epsilon ({\bar \phi}^A) 
\,\equiv \,\epsilon_A\,$,
$\epsilon (\phi^{\ast\,a}_A ) \,=\,\epsilon ( \pi^a_A ) \,=\,
\epsilon_A\,+\,1\,$. 
In this way the ideas of extended BRST quantization in the antifield 
context previously discussed in\cite{BLT1,BLT2,BLT3} are put in a
completely anticanonical setting. 
The extended BRST invariance of the generating
functional, defined on this 6n dimensional space, is equivalent to
the fact that the quantum action $\,W\,$ is a 
solution of the two master equations:
       
\begin{equation}
\label{ME}
{1\over 2} \{ W \,,\, W\,\}^a \,+\,V^a W \,=\,
i\hbar \Delta^a W
\end{equation}

\noindent where the indices $\,a\,=\,1\,,\,2\,$ correspond 
respectively 
to BRST and anti-BRST invariances and the extended form of the 
antibrackets, triangle and $V$ operators read

\begin{equation}
\label{AB}
\{ F\,,\,G\,\}^a\,\equiv \,{\partial^r F\over \partial \phi^A} 
{\partial^l G \over \partial \phi_A^{\ast\,a}}\,+\,
{\partial^r F\over \partial {\bar \phi}^A} 
{\partial^l G \over \partial \pi_A^a\,}
\,-\, {\partial^r F \over \partial \phi_A^{\ast\,a}}\,
{\partial^l G\over \partial \phi^A}\,-\,
{\partial^r F \over \partial \pi_A^a\,}
{\partial^l G\over \partial {\bar \phi}^A} 
\end{equation}
 
\begin{equation}
\Delta^a\,\equiv\, (-1)^{\epsilon_A}\,
 \,{\partial^l \over \partial \phi^A} 
{\partial^l  \over \partial \phi_A^{\ast\,a}}\,+\,
(-1)^{\epsilon_A}\,
{\partial^l \over \partial {\bar \phi}^A} 
{\partial^l  \over \partial \pi_A^a\,}
\end{equation}

\begin{equation}
V^a\,=\,{1\over 2} \epsilon^{ab}\,\Big( \phi_{A b}^{\ast}
{\partial^r \over \partial {\bar \phi}^A}  
- (-1)^{\epsilon_A} \pi_{A\,b} {\partial^r \over 
\partial \phi^A} \Big)\,\,.
\end{equation}

\noindent here and in the rest of the article, 
unless explicitly 
indicated, we are adopting the convention of summing 
over repeated indices.

The Vacuum functional is  normally defined including also an extra 
functional $X$  

\begin{equation}
\label{VAC}
Z\,=\, \int [{\cal D}\phi][\Dc \phi^{\ast}][\Dc \pi]
[\Dc \bar\phi ][\Dc \lambda]\,exp\{{i\over \hbar}\big(
W\,+\,X\big)\}
\end{equation}
 
\ni  that represents gauge fixing and must satisfy 
the  equations

\begin{equation}
{1\over 2} \{ X \,,\, X\,\}^a \,-\,V^a X \,=\,
i\hbar \Delta^a X
\end{equation}

An alternative way of gauge fixing,  using canonical transformations
rather than including the functional $X$ was proposed in \cite{ABG}. We
will use this method in section {\bf 4} for gauge fixing W2 theory. 

Expanding the quantum action in powers of $\hbar $: 
$W\,=\,S\,+\,\hbar M_1\,+ \,...$
we can look at the two first orders of the master equations
 
\begin{eqnarray}
\label{ME2}
 {1\over 2}\{ S \,,\, S\,\}^a &+& V^a S \,=\,0\nonumber\\
  \{ S \,,\, M_1\,\}^a &+& V^a M_1 \,=\,
i \Delta^a S
\end{eqnarray}

For a gauge theory with closed and irreducible algebra, corresponding 
to a classical action $S_0[\phi^i]$, a solution for 
the zero loop order action $S$ is:

\begin{equation}
\label{TAC}
S \,=\, S_0 + \phi_A^{\ast\,a}\,\delta_a \phi^A
+ {1\over 2} {\bar \phi}_A \delta_2 
\delta_1 \phi^A 
\,+\,{1\over 2} \epsilon^{ab} \phi_{A\,a}^{\ast}\,\pi^A_b
\end{equation}

\ni where the $\delta_a $ represent  gauge fixed 
BRST ($a=1$) and anti-BRST ($a=2$) transformations of the fields 
(in other words, for theories with closed algebra, the standard BRST 
extended algebra associated with the gauge theory). In this article we 
will not be dealing with the generalized BRST transformations of the
triplectic formalism\cite{T1,T2,T3} but just with standard transformations
that do not involve the antifields.

Let us consider now the one loop order equation. As it happens in 
the standard BV case, we need to introduce a 
regularization procedure in order to give a well defined 
meaning to the operator $\Delta^a \,S\,$ as there are two 
functional derivatives acting on the same space time point.
If we consider actions of the form (\ref{TAC}) we see that 
the second term in the $\Delta^a\,$ operator will not contribute 
and the important term in the action is just $\phi_A^{\ast\,a}\,
\delta_a \phi^A$. That means we must regularize:

\begin{equation}
 {\partial^l \over \partial \phi^A} 
{\partial^l  \over \partial \phi_A^{\ast\,a}}\,
\Big(  \phi_A^{\ast\,a}\,\delta_a \phi^A \Big)\,\,.
\end{equation}

\noindent If the BRST algebra is such that the BRST and anti-BRST 
transformations are symmetrical, just changing ghosts by antighosts,
then the same regularization can be used in both sectors. Moreover, 
we can use the same regularization used in the standard BV 
quantization.

\section{Extended BRST invariance in W2 gravity}
The classical action corresponding to W2 gravity reads
\cite{DJ}

\begin{equation}
S_0\,=\,{1\over 2\pi} \int d^2x \Big( \partial \phi 
{\bar \partial}
 \phi \,-\,    
h (\partial \phi)^2 \Big) 
\end{equation}

\noindent and the corresponding BRST anti BRST algebra, 
satisfying \break 
$(\delta_1)^2\,=\,
(\delta_2)^2 \,=\,\delta_1 
\delta_2 \,+\,
\delta_2 
\delta_1\,=\,0\,$ is
 
\begin{eqnarray}
\label{Alg1}
\delta_1 \phi &=&  c_{_1} \partial \phi  \no\\
\delta_1 h &=& {\bar \partial} c_{_1} \,-\, h \partial c_{_1} \,
+\,\partial h  
c_{_1} \no\\
\delta_1 c_{_1}  &=&  \partial c_{_1} \,c_{_1}\no\\
\delta_1 c_{_2}  &=& b\nonumber\\
\delta_2 \phi &=&  c_{_2} \partial \phi  \no\\
\delta_2 h &=& {\bar \partial} c_{_2} \,-\, h 
\partial c_{_2} \,+\,
\partial h  c_{_2} 
\no\\
\delta_2 c_{_1}  &=& -b - c_{_2} \partial c_{_1} \,-\,c_{_1}
\partial c_{_2}\no\\
\delta_2 c_{_2}  &=&  \partial c_{_2}\,
c_{_2}\no\\ 
\end{eqnarray}
 
\ni where we are representing BRST and anti-BRST 
transformations 
respectively as 
$\delta_1$ and $\delta_2$.

The general form of the gauge fixed action, after functionally integrating
over the auxiliary fields of the triplectic formalism: $ {\bar \phi}^A $,
 $\,\phi_A^{\ast\,a}\,$ and $\pi_A^{\,a}\,$
is 
 
\begin{equation}
S\,=\,S_0\,+\,\delta_1 \,\delta_2\,B
\end{equation}

\noindent where $B[\,\phi^A\, ]\,$ is a bosonic functional. Therefore the 
ultimate result of triplectic 
quantization would be to build up such an object. However it is   not 
possible to find a bosonic functional $B$ that removes the degeneracy of 
the action $\,S_0\,\,$ using just the  fields of algebra (\ref{Alg1}).
We need more fields in order to obtain such a gauge fixing in W2 theory. 
Inspired in the extended algebra for the bosonic string from ref.
\cite{Ba} we can introduce the bosonic fields $L$ and $\lambda$ and the 
fermionic fields $\eta$ and ${\bar \eta}$ and try transformations of 
the form

\begin{eqnarray}
\label{Alg2}
\delta_1 \eta &=&  
 \partial \eta c_{_1} \,+\, 2 \alpha \partial c_{_1} \eta  \no\\
\delta_1 L &=&  a_1 \eta \,+\, \partial L c_{_1} \,
-\, 2 \alpha \partial c_{_1} L  \no\\
\delta_1 \lambda  &=&   \partial \lambda c_{_1}   
\,-\, 2 \alpha \partial c_{_1} \,\lambda \no\\
\delta_1 {\bar \eta}  &=&  a_2 \lambda \,
+\, \partial {\bar \eta} c_{_1}\,+\,2  \partial c_{_1} {\bar \eta}
\nonumber\\
\delta_2 \eta &=& b_1 \lambda \,+\, 
 \partial \eta c_{_2} \,+\, 2 \alpha \partial c_{_2} \eta  
\no\\
\delta_2 L &=&  b_2 {\bar \eta} \,
+\, \partial L c_{_2} \,-\, 
2 \alpha \partial c_{_2} L
\no\\
\delta_2 \lambda  &=&   \partial \lambda c_{_2}   
\,-\, 2 \alpha \partial c_{_2} \,\lambda \no\\
\delta_2 {\bar \eta}  &=& 
 \partial {\bar \eta} c_{_2}\,+\,
2   \alpha \partial c_{_2} {\bar \eta}
\\ 
\end{eqnarray}

\noindent Extended nilpotency is satisfied if 
$  a_1 b_1 \,+\,  a_2 b_2 \,=\,0\,\,$ for any $\alpha$.
 
\noindent  We will choose $ a_1\,=\, a_2\,=b_1\,=\,-b_2\,=1$ 
and $\alpha\,=\,1$.

In this enlarged space we can choose the gauge fixing boson as

\begin{equation}
\label{boson}
B\,=\,L ( h\,-\,\tilde h )
\end{equation}

\noindent where $\tilde h$ is a BRST anti-BRST invariant background 
field.

\noindent If we redefine the fields as

\begin{eqnarray}
\lambda^\prime &=& -\lambda \,+\, \partial {\bar \eta} c_{_1} \,+\, 
2 \partial c_{_1} {\bar \eta}\,+\,\partial L b \,-\,\partial\eta 
c_{_2}\, -\,\partial^2 L c_{_1} c_{_2} \,-\,
\partial L \partial c_{_1} c_{_2}\,\no\\
&+&
\,\partial^2  c_{_1} L c_{_2} \,+\,
 2 \partial c_{_1} \partial L c_{_2}
- 2 \partial c_2 \eta \,-\,2 \partial c_{_2} \partial L c_{_1}
\,+\,4 \partial c_{_2} \partial c_{_1} L \,-\,
2 \partial b L \no\\
\eta^\prime &=& \eta\,+\, \partial L c_{_1} \,-\,2 \partial c_{_1} L\no\\
b^\prime &=&  {\bar \eta} \,+\,L  \partial c_{_2} 
\end{eqnarray}

\noindent the gauge fixing action gets

\begin{equation}
\label{GB}
\delta_1  \delta_2 
\Big( L ( h\,-\,\tilde h ) \Big)\,=\,
\lambda^\prime ( h\,-\,\tilde h )\,+\, b^\prime 
({\bar \partial} c_{_1} \,-\, h \partial c_{_1} \,+\,\partial h c_{_1} ) 
\,+\,\eta^\prime ({\bar \partial}  \,-\, h \partial  \,
+\,\partial h ) c_{_2} \,+\,
L  ({\bar \partial}  \,-\, h \partial \,+\,\partial h ) b\,\,.
\end{equation}

\bigskip

\noindent The two last terms cancel each other  by a supersymmetric
compensation in the path integral\cite{Ba} while the remaining two 
first terms correspond to the gauge fixing obtained in ref.\cite{DJ} 
in the standard BV scheme with just BRST invariance. 
Thus, the boson $B$ of eq. (\ref{boson}) would appropriately fix the 
gauge of W2 gravity. Now we will see in the next section how to arrive 
at this gauge fixing action of eq. (\ref{GB}) starting from the 
triplectic action.

\section{Gauge fixing by canonical transformations}

One interesting way to get the gauge fixed action of the form 
$S \,=\,S_0\,+\,\delta_1 \,\delta_2\,B$
from the triplectic action (\ref{TAC}) is to perform canonical 
transformations in the triplectic space. These transformations have been
studied in \cite{ABG}. For each of the antibrackets  of eq. (\ref{AB}) 
with $a=1,2$ we introduce a generator 
$F_a\,[\phi^A\,,{\bar \phi}^A\,,\phi^{\ast\,a\,\prime}_A\,,
\pi^{a\,\prime}_A\,\,]\,$ and write out the set of transformations

\begin{eqnarray}
\label{CTR1}
\phi^{A\,\prime} &=& {\partial F_a \over \partial \phi^{\ast\,a\,\prime}_A}
\no\\
\phi_A^{\ast\,a} &=& {\partial F_a \over \partial \phi^A}\no\\
{\bar \phi}^{A\,\prime} &=& {\partial F_a \over \partial \pi^{a\,\prime}_A}
\no\\
\pi_A^{a} &=& {\partial F_a \over \partial {\bar \phi}^A}\,,
\end{eqnarray}

\ni where there is no sum over $a$. If the matrix

\begin{equation}
\label{invers}
 T_a^{\alpha\,\beta} \,=\,{ \partial^r \partial^r  F_a \over 
\partial z^{\ast\,a\,\prime}_\alpha \partial z_\beta }
\end{equation}

\ni (where there is again no sum over $a$ and we are defining 
$\{ z^\alpha \} \,\equiv \, \{ \phi^A \,,\,{\bar \phi}^A \} \,$ and 
$\, \{ z^{\ast \, a\,\prime}_\alpha \} \,\equiv\,
\{ \phi^{\ast\,a\,\prime}_A \,,\,
\pi^{a\,\prime}_A \}\,\,$) is invertible, each of  these transformations, 
for fixed $a=1$ or $2$ will not change  the form of the corresponding 
antibracket.

If additionally the generators $F_1$ , $F_2$  both with non 
singular matrices (\ref{invers})  satisfy also the constraints

\begin{eqnarray}
\label{conditions}
{\partial F_1 \over \partial \phi^{\ast\,1\,\prime}_A} &=&
{\partial F_2 \over \partial \phi^{\ast\,2\,\prime}_A}
\no\\& &\no\\
{\partial F_1 \over \partial \pi^{1\,\prime}_A} &=&
{\partial F_2 \over \partial \pi^{2\,\prime}_A}\,\,.
\end{eqnarray}

\bigskip

\ni Then the complete set of transformations  (\ref{CTR1}) 
including both $a=1$ and  $a=2$ will  leave  the two antibrackets invariant,
preserving the complete triplectic anticanonical structure.
 
The constraints (\ref{conditions}) restrict the possible dependence 
of the generators of these transformations on the variables 
$\phi^{\ast\,a\,\prime}_{A}$ and
$\pi^{a\,\prime}_A\,$. Their general form can be written as 

\begin{equation}
F_a \,=\,{\bf 1}_a\,+\,f_a 
\end{equation}

\ni with

\begin{eqnarray}
{\bf 1}_a &=&  \phi^A \phi^{\ast\,\prime}_{A\,a}
\,+\,{\bar \phi}^A \pi^{\prime}_{A\,a} \no\\& &\no\\
f_1 &=& g_1 [\phi\,,{\bar \phi}]\,+
\,g_3^A[\phi\,,{\bar \phi}] \pi^{ 1\,\prime}_A \,+\,
g_4^A [\phi\,,{\bar \phi}]\phi^{\ast\,1\,\prime}_A \no\\& &\no\\
f_2 &=&  g_2[\phi\,,{\bar \phi}] 
\,+\, g_3^A [\phi\,,{\bar \phi}] \pi^{2\,\prime}_A\,+\,
g_4^A [\phi\,,{\bar \phi}]  \phi^{\ast\,2\,\prime}_A \,\,,
\end{eqnarray}

\ni where we have  explicitly separated an identity operator 
${\bf 1}_a\,$  just for future convenience.

\bigskip
Now going back to the equation (\ref{CTR1}) we see that general  triplectic 
canonical transformations can be put in the form

\begin{eqnarray}
\label{CTR2}
\phi^{\prime\,A} &=& \phi^{A}\,+\,g^[\phi\,,{\bar \phi}]
\no\\
\phi^{\ast\,a\,}_A &=& \phi^{\ast\,a\,\prime}_A \,+\,
{\partial^r g_a \over \partial \phi^A}[\phi\,,{\bar \phi}]
\,+\,{\partial^r g_3^B \over \partial \phi^A} [\phi\,,{\bar \phi}] 
\pi_B^{a\,\prime}
\,+\,{\partial^r g_4^B \over \partial \phi^A}[\phi\,,{\bar \phi}]
\phi_B^{\ast\,a\,\prime}  \no\\
{\bar \phi}^{A\,\prime} &=& {\bar \phi}^A \,+\, 
g_3^A [\phi\,,{\bar \phi}]\no\\
\pi_A^{a} &=& \pi_A^{a\,\prime} \,+\,
 {\partial^r g_a \over \partial {\bar \phi}^A}[\phi\,,{\bar \phi}]
\,+\,{\partial^r g_3^B \over \partial {\bar \phi}^A}
[\phi\,,{\bar \phi}] \pi_B^{a\,\prime}
\,+\,{\partial^r g_4^B \over \partial {\bar \phi}^A} [\phi\,,{\bar \phi}]
\phi_B^{\ast\,a\,\prime}  \no\\
\end{eqnarray}

The condition that a canonical transformation reproduces the gauge 
fixing corresponding to some boson B, after we express the result 
in terms of the transformed fields and impose the condition that  
$ {\bar \phi}^{A\,\prime} $, $\,\phi_A^{\ast\,a\,\prime}\,$
and $\pi_A^{\,a\,\prime}\,$ are set to zero reads

\begin{equation}
\label{general}
{\partial f^{\prime}_a\over \partial \phi^A }\delta_a 
\phi^A \,+\,{1\over 2} g^{\prime}_3 \delta_2 \,
\delta_1 \phi^A \,-\,
{1\over 2}\epsilon^{ab} {\partial f^{\prime}_a\over \partial \phi^A }
{\partial f^{\prime}_b\over \partial {\bar \phi}^A }\,\,=\,
 \delta_2 \delta_1 B [\phi^A]\,\,,
\end{equation}

\ni where we are defining the primed functions as the corresponding 
function, written in terms of $\phi^A$ and 
$ {\bar \phi}^{A\,\prime} $, taken at $ {\bar \phi}^{A\,\prime}\,=\,0 $

\begin{equation}
f^{\prime}_a [\phi]\,=\, 
f_a [\,\phi\,,\,{\bar \phi}(\phi\,,
\,{\bar \phi}^{\prime}\,)\, 
]\vert_{_{{\bar \phi}^{\prime}\,=\,0}}
\end{equation} 

\ni and a similar definition for $\,g^{\prime}_i\,$.

Considering our W2 case, two  illustrative  possibilities
are to choose

\begin{eqnarray}
g_1 &=& g_3 \,=\, g_4 \,=\, 0\nonumber\\
g_2 &=& \delta_1 \,\Big( L ( h - {\tilde h}) \Big)
\,=\,L ({\bar \partial} c_{_1} \,-\,h \partial c_{_1} \,+\,\partial h 
c_{_1} \,) \,+\, (\eta\,+\,\partial L c_{_1} \,-\,2\partial l 
c_{_1}\,)(h - {\tilde h}),
\end{eqnarray}

\noindent or

\begin{eqnarray}
g_1 &=& -\,\delta_2 \,\Big( L ( h - {\tilde h)}\Big)
\,=\,L ({\bar \partial} c_{_2} \,-\,h \partial c_{_2} \,+\,
\partial h c_{_2} \,) \,+
\, (- {\bar \eta}\,+\,\partial L c_{_2} \,-\,2\partial l c_{_2}\,)
(h - {\tilde h}) \nonumber\\
g_2 &=& g_3 \,=\, g_4 \,= \, 0
\end{eqnarray}

\noindent In both cases we get the gauge fixing action (\ref{GB}) if we 
perform the corresponding transformation in the fields of action $S$ and 
then set all the primed antifields to zero.

\section{One loop order}
The first point that must be investigated is the possible effect of 
the introduction of the fields  $L$ , $\lambda$ , $\eta$ and 
${\bar \eta}$ in the anomalies of the model. 
In other words, we must see if the cohomology of our extended formulation
is the same as that from the original one.
We can introduce a filtration\cite{PS}
${\cal N}$ that counts the number of fields and expand the BRST anti-BRST 
operators according to this filtration 
$ \delta_1\,=\, \delta_1^{(0)}\,+\,
\delta_1^{(1)}\,\,$;
$\,\, \delta_2\,=\,\delta_2^{(0)}\,+\,
\delta_2^{(1)}\,$. The first order piece of the algebra reads

\begin{eqnarray}
\label{Alg21}
\delta_1^{(0)} \eta &=&  0  \no\\
\delta_1^{(0)} L &=&  \eta  \no\\
\delta_1^{(0)} \lambda  &=& 0 \no\\
\delta_1^{(0)} {\bar \eta}  &=&  \lambda 
\nonumber\\
\delta_2^{(0)} \eta &=& \lambda   \no\\
\delta_2^{(0)} L &=&  -  {\bar \eta} \no\\
\delta_2^{(0)} \lambda  &=&  0  \no\\
\delta_2^{(0)} {\bar \eta}  &=& 0\,\,\,. \\ 
\end{eqnarray}

Looking at this algebra we realize that this  fields  form doublets
with respect to both the BRST and anti-BRST transformations.
By a doublet one  means a pair of fields say $u,v$ whose transformations
are of the form $\delta u\,=\,v\,,\,\delta v\,=\,0\,$. 
Fields that show up just in doublets are absent from the cohomology of the 
BRST operator (or correspondingly homology of the anti BRST operator). 
Moreover, the cohomology ($a=1$) or homology ($a=2$) of the operator
$\delta_a$ is contained in the cohomology (homology)  
of the corresponding
$\,\delta_a^{(0)}\,$ , $\,a\,=\,1\,$.  
Thus we conclude that the inclusion of the fields $L\,,\,\lambda \,  
,\,\eta\,$ and $\, {\bar \eta}$ does not change the cohomology 
(homology)of the W2 theory.  We can then consider the same quantum 
correction $\Delta S$ to the
first order master equations as for the standard formulation of W2.
The calculation of $\Delta S$ depends on the introduction of a 
regularization procedure and the result depends on the result. 
But all the possible results differ just by trivial terms (in the 
cohomological sense). The simplest way to right the results of \cite{JST}
adapting them to the extended symmetry is:

\begin{eqnarray}
\label{DeltaS} 
(\Delta^1 S\,)_{Reg}
 &=& \,-\, {1\over 12\pi}
 \int d^2x 
 \Big(   c_{_1} \, \partial^3 h \Big)
\nonumber\\
(\Delta^2 S\,)_{Reg}
 &=& \,-\, {1\over 12\pi}
 \int d^2x 
 \Big(   c_{_2} \, \partial^3 h \Big)
\end{eqnarray}

The presence of this term in the master equation at one loop order means a 
breaking in the BRST invariance. What one normally does in the BV 
quantization is then to introduce an extra Wess Zumino field $\theta$ 
representing the extra degree of freedom corresponding to the anomalous 
breaking of gauge invariance. The BRST extended version of this procedure
 would correspond to define this new field with the transformations:

\begin{eqnarray}
\delta_1 \theta &=& \partial c_{_1}\,+\,c_{_1} \partial \theta
\nonumber\\
\delta_2 \, \theta &=& \partial c_{_2} \,+\, c_{_2}\partial 
\theta\,\,.
\end{eqnarray}

\noindent Then we verify that the counterterm 

\begin{equation}
M_1 \,=\,{1\over 24\pi} \int d^2x \Big( \partial \theta 
{\bar \partial} \theta \,-\, h \partial \theta \partial \theta\,+\,
h \partial^2 \theta \Big) 
\end{equation}

\noindent solves the master equations:

\begin{equation}
\{ S \,,\, M_1\,\}^a \,+\, V^a M_1 \,=\,
i (\Delta^a  S \,)_{Reg}
\end{equation}

\noindent for both $\,a\,=\,1,2\,$. As a remark we mention that a 
different approach could be taken to the addition
of the field $\theta$ to the theory. 
We mean, one could  include another ghost associated with the invariance of 
the classical action with respect to any transformations in $\theta\,$.
As a result the master equation would never be solved. One would only be 
able to shift the anomaly to this new symmetry\cite{JST}, leaving the 
original gauge symmetry unbroken, but not the BRST 
(and anti-BRST symmetries). 
We will keep here the point of view of \cite{BM} that the field 
$\theta $ represents the new degree of freedom that shows up at 
quantum level as a consequence of the  anomalous breaking of the 
original gauge symmetry. Thus we do not add any extra ghost.

Then we see that at one loop order the triplectic quantization reproduces 
the so called Wess Zumino mechanism of restoring the gauge invariance 
of an anomalous gauge theory by means of the introduction of an extra 
degree of freedom.

\section{Conclusion}
We have seen that, enlarging the space of fields, it is possible to 
formulate  W2 gravity with extended BRST invariance. We have proven 
that these enlargement of the representation do not change the cohomology 
of the theory.
We have calculated the one loop order corrections in the triplectic 
quantization for the model. We have seen that the anomalies and 
counterterms of the BRST and anti BRST sectors are essentially the same,
up to changing ghosts by antighosts.
The question that can then be raised then is: how  general is this result?
In the present case this  happens because the BRST and anti BRST algebras
are symmetric and thus one  trivially concludes that the cohomology 
(homology) is the same for both symmetries (again, up to changing 
ghosts by antighosts).
It seems an interesting future task to look, in some gauge
theory, for a $\delta_2$ symmetry satisfying the extended algebra 
$(\delta_2)^2 \,=\,\delta_1 
\delta_2 \,+\,
\delta_2 
\delta_1\,=\,0\,$ but with a different homology.

 Acknowledgements: The authors are partially supported by CNPq., FINEP and 
FUJB (Brazilian Research Agencies).
\vfill\eject

\end{document}